# Discrete Solitons of the Ginzburg-Landau Equation

**Mario Salerno[1] and Fatkhulla Kh. Abdullaev[2]**

[1] Dipartimento di Fisica "E.R. Caianiello", and INFN, Gruppo Collegato di Salerno, Università di Salerno, Via Giovanni Paolo II, 84084 Fisciano, Salerno, Italy,

[2] Physical-Technical Institute, Uzbek Academy of Sciences,100084, Tashkent, Uzbekistan

**Abstract**  In this chapter we review recent results concerning localized and extended dissipative solutions of the discrete complex Ginzburg-Landau equation. In particular, we discuss discrete diffraction effects arising both from linear and nonlinear properties, the existence of self-localized dissipative solitons in the presence of cubic-quintic terms and modulational instability induced by saturable nonlinearities. Dynamical stability properties of localized and extended dissipative discrete solitons are also discussed.

## I. Introduction

The complex Ginzburg-Landau equation (GLE) is a fundamental equation of physics which appears in many different contexts, including phase transitions, superconductivity, nonlinear optics, non-equilibrium fluid dynamics, Bose-Einstein condensates, etc. [1–5]. Being intrinsically nonlinear the GLE displays a very reach dynamical range of behaviors ranging from chaotic motion to regular structures as vortex and dissipative solitons.

Properties of dissipative solitons of the continuous GLE equation have been largely investigated in the past decades [6]. They were predicted by Pereira-Stenflo in [7] (see also [8]), where the exact soliton solution of the one-dimensional cubic Ginzburg-Landau equation with filter was obtained for the first time. In contrast with solitons of Hamiltonian systems that usually appear as one-parameter families of solutions and whose existence relies on the balance between nonlinearity and dispersion, the existence of dissipative solitons of GLE requires two conditions to be simultaneously satisfied e.g. the equilibrium between nonlinearity and dispersion and the equilibrium between dissipation and amplification. Moreover, dissipative solitons do not appear in families but exist just for specific values of the parameters appearing in the equation. Dissipative solitons and breathers can also appear in the nonlinear Schrödinger equation in the presence of different types of complex periodic potentials [9].

Quite recently a great deal of attention has been devoted also to the study of nonlinear discrete optical systems modeled by a discrete version of the GLE (DGLE) [10–18], which is similar to the discrete nonlinear Schrödinger equation (DNLSE) but with dissipative and amplification effects included. The DGLE has been used to describe a number of physical systems, including arrays of waveguides with amplification and damping, arrays of semiconductor lasers [19], arrays of exciton-polariton condensates [20], frustrated vortices in hydrodynamics [21], dissipative discrete nonlinear electrical lattices with nearest-neighbor interaction [22], etc. In particular, in optics the DGLE appears in problems of beam propagation in the array of the nonlinear optical waveguides with Kerr and resonance nonlinearities [23–25]. The Kerr medium is assumed active and with intrinsic, saturable gain and damping. The existence of a variety of nonlinear localized modes in these systems, including moving discrete dissipative breather-solitons [23] and vortex dissipative solitons [26], was demonstrated.

Similarly to the continuous case, discrete solitons require the balance both of dispersion and nonlinearity and of dissipation and amplification. In this context discrete dissipative solitons have been investigated for power law [13] and saturable [18] nonlinearities, for complex extensions of the Ablowitz-Ladik equation [14, 27], for DNLS-type equations with cubic-quintic nonlinearities [28–31].

In the case of saturable nonlinearities, the study of solitons has been restricted mainly to the conservative case. In particular, discrete solitons of DNLSE with saturable nonlinearity were investigated in Refs. [32, 33] and discrete breathers for the same type of equation in Ref. [34]. In spite of the relevance of this type of nonlinearity for optics, existence and stability of dissipative solitons in the presence of a saturable nonlinearity are poorly investigated. In the continuum case saturable nonlinearities have been considered in the one-dimensional complex Ginzburg-Landau equation both for scalar and vectorial cases [35]. Modulational instability and stopping of Kerr self-focusing induced by nonconservative effects have also been investigated in the multidimensional continuous complex GL type equation with nonlinear saturation [36, 37].

In this chapter we review some recent results on localized and extended dissipative solutions of the discrete complex Ginzburg-Landau equation. In particular, we discuss discrete diffraction effects arising both from linear and nonlinear properties, the existence of self-localized dissipative solitons in the presence of cubic-quintic terms. we discuss the existence and stability of dissipative solitons of the DGLE with saturable nonlinearity. In this last case we consider the problem of instability of nonlinear plane waves solution under weak modulations, e.g. the modulational instability (MI) problem, which allows to define the region of parameters where solitons and train of solitons can be formed and to construct discrete dissipative solitons and nonlinear periodic waves of DGLE with saturable nonlinearity. In particular, we provide explicit analytic expressions for periodic dissipative solitons solutions in the form of elliptic functions both on zero and on a finite background. The stability properties of these solutions are investigated by means of direct numerical simulations of their propagation under the DGLE. As a result, we show that while discrete periodic waves and solitons on a zero background are stable, they become modulationally unstable on a finite background. Effects of a linear ramp potential on stable localized dissipative solitons will be also briefly considered.

## II. The model and linear dispersion relation

The cubic-quintic DGLE can be obtained from a corresponding space-periodic continuous model in the tight binding approximation [38], and in normalized units can be written in full generality as [39]

$$i\frac{d\psi_n}{dz} = -\Gamma(\psi_{n+1} + \psi_{n-1}) + i\gamma_1\psi_n + (\gamma_3 + \gamma_5|\psi_n|^2)|\psi_n|^2\psi_n. \quad (1)$$

with $\gamma_1$ real and $\Gamma = \Gamma_R + i\Gamma_I$, $\gamma_3 = \gamma_{3R} + i\gamma_{3I}$, $\gamma_5 = \gamma_{5R} + i\gamma_{5I}$ complex parameters. In the optical context Eq. (1) arises in connection with semiconductor laser arrays and optical amplifiers [39, 40] but it can be also used to model mean field properties of open Bose-Einstein condensates (BEC) trapped in deep optical lattices [41, 42]. In the optical context (resp. in the BEC context) $z$ denotes the propagation length (resp. the time), $\psi_n$ the amplitude of the electromagnetic wave on site $n$ (resp. the condensate wavefunction at site $n$), $\Gamma_R$ and $\Gamma_I$ are the real and imaginary parts of the complex discrete diffraction in the paraxial approximation (resp. inter-well tunneling constant in BEC), $\gamma_{3R}, \gamma_{5R}$ denote the strengths of the cubic and quintic nonlinearity (two and three body interaction terms in BEC) while $\gamma_1, \gamma_{3I}, \gamma_{5I}$ are real coefficients related to the gain/loss mechanisms present in the optical system (resp. in open BEC). In the following we adopt the notation appropriate for the optical context.

To investigate nonlinear property of this equation it is convenient to start the band structure of the underlying linear system, this being important, at least for not too big nonlinearities, to understand the types of localized dissipative structures one can have in the system. For this we set $\gamma_3 = \gamma_5 = 0$ in Eq. (1) and look for plane wave solutions of the type $\psi_n = \exp i(kn - \Omega z)$ with $k \in R$ denoting the lattice quasi-momentum and $\Omega$ complex, i.e. $\Omega \equiv \Omega_R + i\Omega_I$, denoting the propagation wave number (in nonlinear Schrödinger lattices $z$ plays the role of time and $\Omega$ corresponds to the complex energy, in $\hbar$ units). Direct substitution of the above wave in Eq. (1) leads to the following dispersion relations:

$$\Omega_R = 2\Gamma_R\cos(k), \quad \Omega_I = \gamma_1 - 2\Gamma_I\cos(k), \tag{2}$$

and the periodicity of the lattice (lattice constant being fixed to 1) permits to restrict the resulting real and imaginary bands to the first BZ, e.g. $k \in [-\pi, \pi]$. In Fig.1 are shown typical linear bands for different $\gamma_1, \Gamma$ parameters.

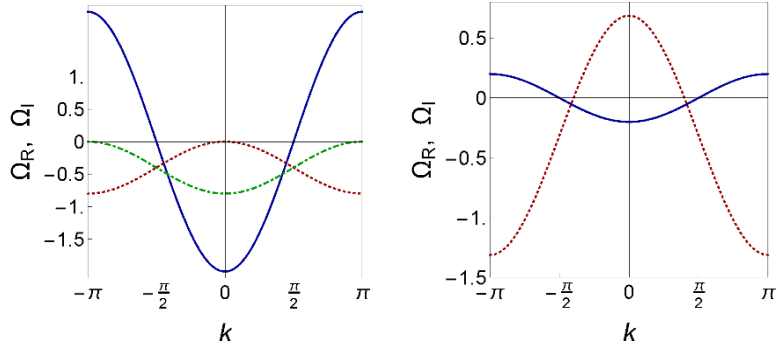

**Fig. 1.** Typical real (continuous blue curves) and imaginary (dotted red curves) linear bands in Eq. (2). Parameter values are fixed as $\Gamma_R = 1.0, \Gamma_I = -0.2, \gamma_1 = -0.4$, in the left panel and as $\Gamma_R = 0.1, \Gamma_I = -0.5, \gamma_1 = -0.3123246628$ in the right panel. The dot-dashed green curve in the left panel refers to the imaginary band obtained for $\Gamma_I = 0.2$, with real band and other and the same.

While the real band has a direct correspondence in closed nonlinear lattices, the imaginary band is typical of open systems, playing an important role for stability of the stationary solutions in the system. Thus, for example, from Eq. (2) it follows that any zero amplitude solutions in the Brillouin zone (BZ) can be stable if $\Omega_I(k) < 0$ for $k$, this implying that $\gamma_1 \leq 2\Gamma_I$. On the other hand stationary Bloch states of a given $k$ can remains stable under z-propagation only if the corresponding $\Omega(k)$ is exactly zero. As example, in the left panel of Fig.1 we shown band structures for the cases in which $\Omega_I$ becomes zero at $k = 0$ (red dotted curve) or at the edges of the BZ $k = \pm\pi$ (green dot-dashed curve) corresponding to the cases $\Gamma_I = -0.2$ and $\Gamma_I = 0.2$, respectively. Notice that the other parameters are fixed the same and so is the $\Omega_R$ band for the two cases.

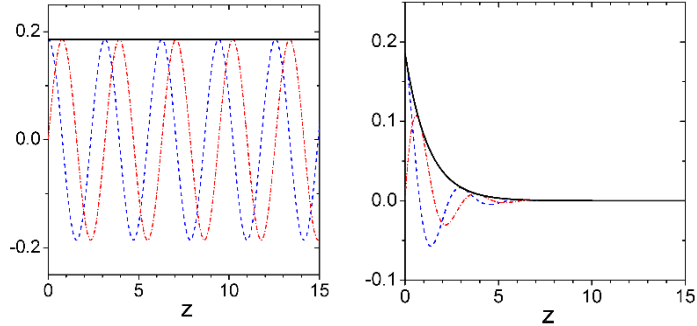

**Fig. 2.** Numerical z-evolution of the uniform Bloch state of the band structure depicted in the left panel of Fig.1 at $k = 0$, for parameters values $\Gamma_I = -0.2$ (left panel) and $\Gamma_I = 0.2$ (right panel). Other parameters are fixed as in the left panel of Fig.1. In the figures are depicted the modulo $|\psi_n|$ (black continuous curve), the real (dashed blue curve) and imaginary (red dot-dashed curve) parts of $\psi_n$ at a generic site $n$ of the lattice.

Thus, the ground state of the $\Omega_R$ band (i.e. the Bloch state at $k = 0$) is stable in the first case (e.g. $\Gamma_I = -0.2$) but not for second case (e.g. $\Gamma_I = 0.2$), as one can see from Fig.2 where the z-propagation for the uniform ground state $\psi_n = 0.185695, \forall n$ are reported for both cases. Obviously the opposite will be true for the highest excited Bloch state $k = \pi$ (not shown here for brevity). All other k-Bloch states in the bands depicted in Fig.1 will either grow, if their $\Omega_I(k) > 0$, or decay into the zero amplitude background if their $\Omega_I(k) < 0$. From this it is clear that the balance between dissipation and gain is very crucial for the existence of stable solutions.

## III. Dissipative Solitons of the DGLE

In analogy with the nonlinear Schrödinger lattices one can expect that dissipative localized states occur from instabilities of the Bloch states either at the center or at the edges of the BZ, depending on the signs of the nonlinearities. For the NLS equation optical lattices this was indeed proved in the small amplitude limit, starting from exact stable Bloch states of the underlying linear problem and using perturbation theory. Moreover, the mechanism for the creation of nonlinear excitations in the band gaps was ascribed to the MI of Bloch states [43]. In the complex case this is complicated by the further requirement of the gain/loss balance needed for stationarity but in principle it is the same. This means that nonlinear wave-packet will bifurcate for the edges of the bands edges and moving into the semi-infinite lower and upper gap as nonlinearity is increased. In the small amplitude limit, as superposition of k-Bloch states with $k$ values centered around either $k = 0$ or $k = \pm\pi i$, depending on the signs of $\gamma_{3R}, \gamma_{5R}$, with corresponding imaginary band values $\Omega_I(k)$ appropriate for stability. The MI will be discussed in more detail in the next section for the case of saturable nonlinearity.

Although it is possible to obtain for some specific case exact solutions of the DGLE, in general one must recourse to numerical methods. This can be done by substitution the stationary ansatz $\psi_n(z) = (u_n + iv_n)\exp(-i\Omega z)$ into Eq. (1) and solving the resulting algebraic system for $u_n, v_n$ either by Newton iterations or by self-consistent diagonalizations of the non-hermitian and nonlinear eigenvalue problem [44]. Typical examples of dissipative discrete solitons obtained numerically are depicted in Figs.3 and 4. Notice that the band structure of the underlying linear problem corresponding to the unstable intrasite symmetric dissipative soliton reported in the left panels of Fig.4 was depicted in the left panel of Fig.1. From this figure we see that the soliton instability under the z-propagation correlates with the existence of a wide k-interval around $k = 0$ for which $\Omega_I(k) > 0$ and corresponding linear Bloch states, unstable.

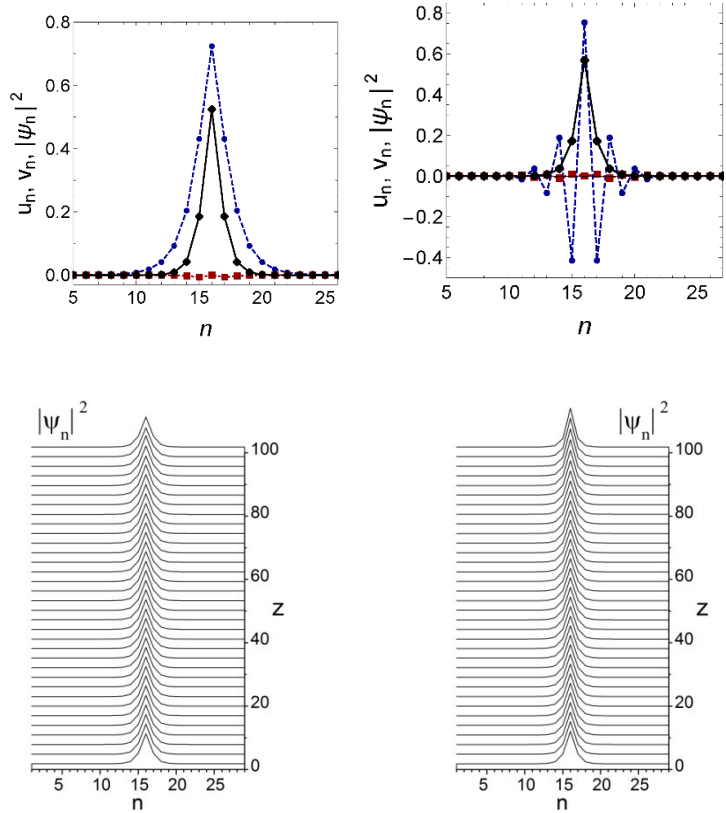

**Fig. 3.** Typical onsite-symmetric dissipative solitons (top panels) and their corresponding z-propagation (bottom panels) obtained from direct numerical diagonalizations/integrations of Eq. (1). The soliton in the top left panel has $k_R = -2.68251$ located below the linear real band at the center ($\theta = 0$) of the BZ. The corresponding z-propagation is depicted in the bottom left panel. Parameter values for this case are fixed as: $\Gamma = 1.0 + 0i, \gamma_1 = -0.1, \gamma_3 = -2.847313757 + 0.3i, \gamma_5 = 0. - 0.6i$. The soliton in the top right panel has $k_R = 2.72421$ located above the linear real band at the edge ($\theta = \pi$) of the BZ. The corresponding z-propagation is shown in the bottom right panel. Parameter values for this case are fixed as: $\Gamma = 1.0 - 0.01i, \gamma_1 = -0.447102453, \gamma_3 = 2.52 + 0.3i, \gamma_5 = 0.6 - 0.28i$. In top panels the continuous black lines with diamonds, blue dashed lines with dots, and red dotted lines with squares, correspond to the modulo square, and real and imaginary parts of $\psi_n$, respectively.

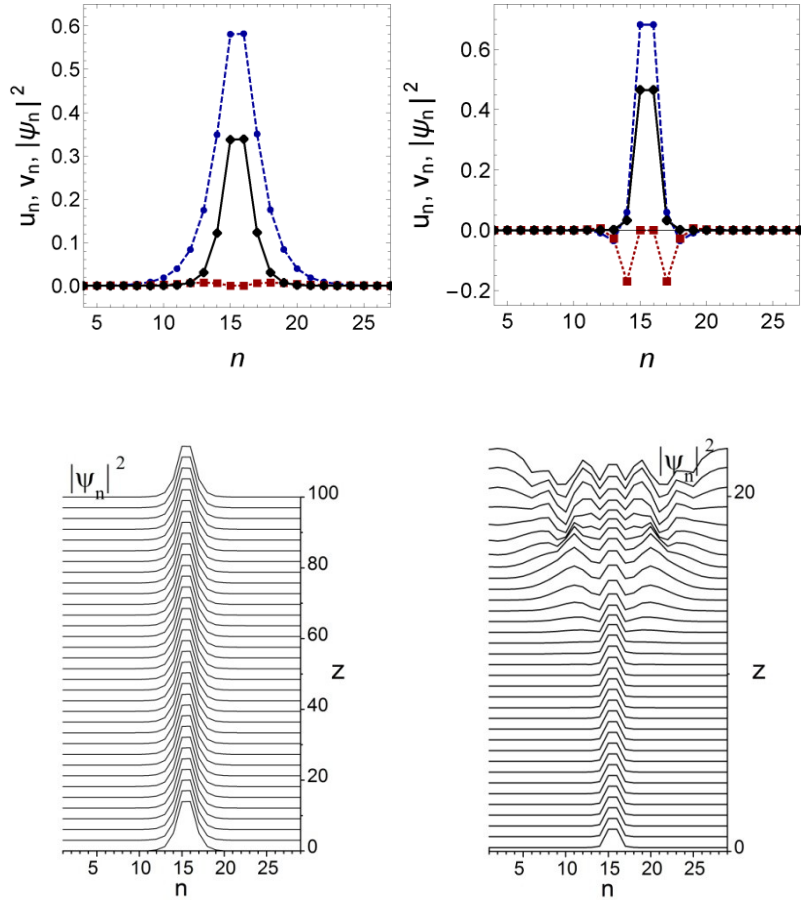

**Fig. 4.** Same as in Fig.3 but for inter-site symmetric dissipative solitons obtained from direct numerical diagonalizations/integrations of Eq. (1). Parameter are fixed for the left panels as: $\Gamma = 1.0 + 0i, \gamma_1 = -0.4573, \gamma_3 = -3.8 + 0.3i, \gamma_5 = 2.6 - 0.4i$, and for the right panels as $\Gamma = 0.1 - 0.5i, \gamma_1 = -0.312325, \gamma_3 = -4.0 - 0.6i0.3i, \gamma_5 = -0.5 - 0.1i$. In both cases the solitons have real part of the propagation wave-number below the bottom edge of the linear $k_R$ band at $\theta = 0$.

## IV. Saturable nonlinearity and MI analysis

As is well known, the MI is a fundamental dynamical phenomenon responsible for soliton and pattern generation in nonlinear systems [45]. In comparison to the continuous (non periodic) case, the MI in the discrete case displays novel properties since the discrete diffraction makes possible to have MI also for defocusing (e.g. repulsive) nonlinearity. Since the parameter region where plane waves become modulationally unstable coincides with the existence region of solitons, one has that nonlinear lattices can support solitons also for defocusing interactions. This fact is true also in the continuous case, if a periodic potential is present [43]. Different discretizations of the same continuous nonlinearity can have different effects on the MI and, correspondingly, can lead to different conditions for the existence of soliton solutions [46]. We also remark that in the DNLSE for small wavenumbers of the nonlinear plane wave, all modulations become unstable if the power excess a threshold value [34]. For DNLSE with a saturable nonlinearity the gain and critical frequency are decreased in comparison with the Kerr nonlinearity model [29]. Experimentally discrete MI has been observed in the array of nonlinear optical waveguides [47] and in photovoltaic crystals [48].

In this and in the next section we study MI and exact solutions of DGLE with saturable nonlinearity and saturable gain/loss which appear when one considers the propagation of beams in the array of nonlinear optical waveguides with active Kerr medium and resonant interaction. The general form of the model for this was introduced in Refs. [23, 24] as

$$i\frac{d\psi_n}{dz} + \Gamma(\psi_{n+1} + \psi_{n-1}) - [|\psi_n|^2 - if_d(|\psi_n|^2)]\psi_n + Qf_{is}(|\psi_n|) = 0. \quad (3)$$

where $f_d(x)$, $x \equiv |\psi_n|^2$, is a real function that describes the amplification and absorbtion of each waveguide of the array of the form [24, 25, 49]

$$f_d(x) = -\delta + \frac{g}{1+x} - \frac{a}{1+bx}, \quad (4)$$

while the function $f_{is}(x)$ and the strength $Q$ are related to the inter-site Kerr nonlinear refractive index (see [23, 24] for details). The parameters $\delta$ describe the linear non resonant losses while $g$, $a$ are the strengths of the saturable gain and absorption, respectively, and b is the ratio between the gain and absorption saturation intensities. For simplicity in the following we restrict to the case $Q = 0$, $b = 1$ and rewrite Eq. (3) in the form

$$i\frac{d\psi_n}{dz} = -\Gamma(\psi_{n+1} + \psi_{n-1}) + i\gamma_1\psi_n + \gamma_3 \frac{|\psi_n|^2}{1+\mu|\psi_n|^2}\psi_n, \qquad (5)$$

where we replaced the Kerr nonlinearity with the saturable nonlinearity, and denoted $\gamma_1 = \delta$ and $\mu$ the parameter controlling the nonlinearity saturation (notice that while in Eq. (3) the field $\psi_n$ is normalized according to the gain saturation intensity and the strength while in Eq. (5) the nonlinearity saturation has been made explicit). Moreover, without loss of generality, we fix $\Gamma = 1 - i\alpha$, $\gamma_3 = -\nu + i\gamma$.

To study the MI we notice that Eq. (5) supports nonlinear plane wave solutions of the form $\psi_n = A\exp(i(kn - \omega t))$, with amplitude $A$, wave numbers $k$ and frequency $\omega$ satisfying the following nonlinear dispersion relation:

$$A^2 = -\frac{\delta + 2\alpha\cos(k)}{\delta\mu + \gamma + 2\mu\alpha\cos(k)}, \quad k \neq \pm\arccos\left(\frac{\delta\mu+\gamma}{2\mu\alpha}\right) \pm \pi,$$

$$\omega = 2\cos(k)\left(\frac{\nu\alpha}{\gamma} - 1\right) + \frac{\nu\delta}{\gamma}. \qquad (6)$$

For $\omega > 0$ there are two possibilities for plane wave existence, e.g. i) for $\nu/\mu > 0$, the frequency must vary in the interval $2 - \nu/\mu < \omega < 2$; ii) for $\nu/\mu < 0$, the frequency must vary in the interval $2 < \omega < 2 + |\nu/\mu|$. In the case $\omega < 0$ we find that the frequency must be varied in the interval $2 - \nu/\mu < \omega < 0$, with $\nu/\mu > 2$. Taking into account that $\omega$ is defined by Eq. (6), one can easily derive restriction on parameters for the existence of plane waves at special points of k-space: at $k = 0$ (unstaggered solution) and at $k = \pi$ (staggered solution). For the staggered solution we find the restriction $\delta \leq 2\alpha$, while for the unstaggered $k = 0$ solution we find that $\nu(2\alpha + \delta)/\gamma \leq 4$ must be satisfied.

To analyze MI in Eq. (5) we look for solutions of the form

$$\phi_n(z) = (A + \phi_n(z))\exp(i(kn - \omega z)), \phi \ll A. \qquad (7)$$

By substituting into Eq. (5) we get

$$i\phi_{n,t} + (1 - i\alpha)(\phi_{n+1}e^{ik} + \phi_{n-1}e^{-ik} - 2\cos(k)\phi_n) +$$

$$(\nu - i\gamma)\frac{A^2}{(1+\mu A^2)^2}(\phi_n + \phi_n^*) = 0. \qquad (8)$$

By looking for solutions of Eq. (8) of the form

$$\phi_n = Be^{i(Qn-\Omega t)} + C^*e^{-i(Qn-\Omega^* t)}, \qquad (9)$$

with $B, C, \Omega$ complex numbers, one readily obtains the following dispersion relation is obtained

$$\Omega^2 - \Lambda_1 \Omega - \Lambda_2 = 0, \tag{10}$$

Where

$$\Lambda_1 = 2\big(2S + i(2\alpha\Delta + \gamma D)\big),$$
$$\Lambda_2 = 4(1+\alpha^2)(\Delta^2 - S^2) + 4D\Delta(\nu + \alpha\gamma) + 4i(\alpha\nu - \gamma)DS, \tag{11}$$

and

$$\Delta = \cos(k)(\cos(Q) - 1), \quad D = \frac{A^2}{(1+\mu A^2)^2}, \quad S = \sin(k)\sin(Q).$$

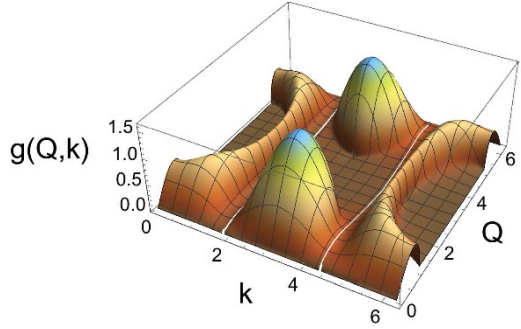

**Fig. 5.** MI gain $g(Q, k)$ in Eq. (12) versus wavenumbers $Q, k$ for parameter values $\nu = 3, \mu = 1, \alpha = 0.01/3, \delta = -0.01, \gamma = 0.012$.

From these equations the MI gain $g(Q, k) = |\text{Im}(\Omega(Q,k)|)$, is readily obtained as

$$g(Q,k) = |(2\alpha\Delta + \gamma D) + \frac{1}{\sqrt{2}}\sqrt{-F + \sqrt{G^2 + F^2}}|, \tag{12}$$

with the functions $F, G$ given by

$$F = (4S^2 - (2\alpha\Delta + \gamma D)^2) + 4(1+\alpha^2)(\Delta^2 - S^2) + 4D\Delta(\nu + \alpha\gamma),$$
$$G = 4S\alpha(2\Delta + \nu D). \tag{13}$$

In Fig.5 we show typical dependence of the MI gain on wavevectors $Q, k$ for the focusing case. Notice that the white open regions visible in the figures correspond to the lines $k = \pm\cos^{-1}\left(\frac{\gamma+\delta\mu}{2\alpha\mu}\right) + \pi$ on which the wavevector $k$ is not defined (see Eq. (6) ). From this analysis the existence of nonlinear local-ized and extended solutions of Eq. (5) is expected. In the next section we shall confirm the existence of dissipative solitons and cnoidal wave solutions by providing few exact solutions and by investigating their stability by means of numerical integrations.

## V. Exact dissipative discrete soliton solutions

Exact dissipative discrete soliton solutions of different types were obtained in [18] by assuming specific ansatzes that involve elliptic functions [50]. Few of them are listed below.

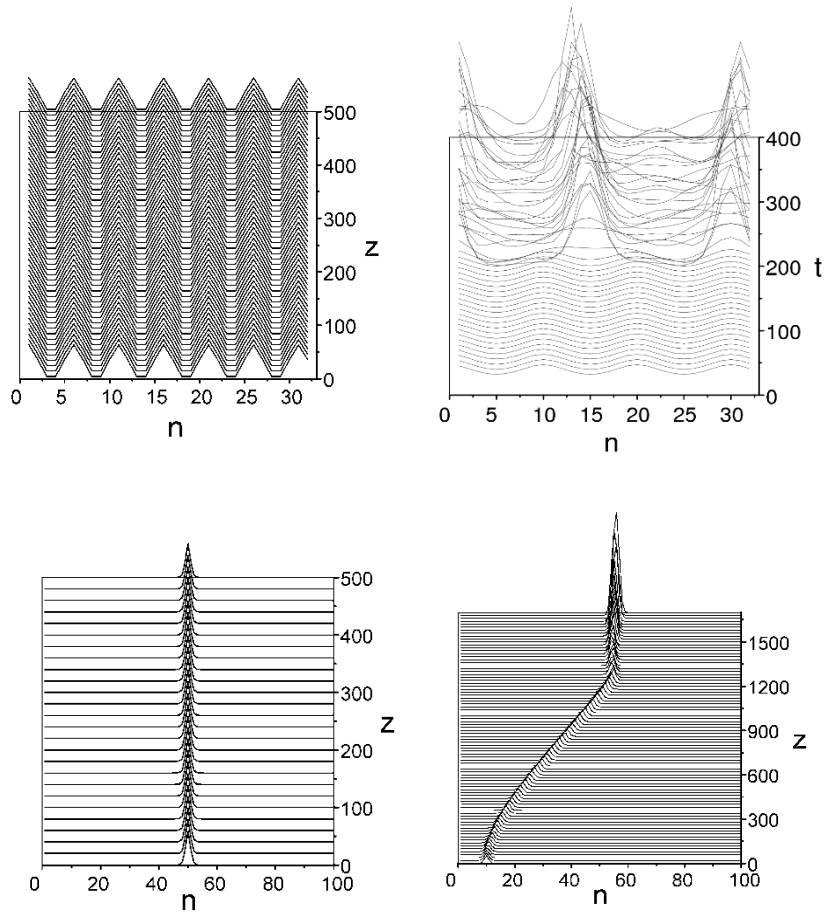

**Fig. 6. Top panels**. Z-propagation of the modulo square of the periodic dissipative soliton trains in Eq. (20) (top left) and in Eq. (17) (top right) as obtained from direct numerical integration of Eq. (5). Parameter values are fixed as in Eq. (22) with $\alpha = 0.001, m = 0.5, \beta = 2K(m)/N_p$ for the top left panel and as in Eq. (19) with $\alpha = 0.01, m = 0.32, \beta = 4K(m)/N_p$ for the top right panel. In both cases the number of lattice points per period is $N_p = 10$ and the total number of points along the line is 30. The cn-solution remain stable and dn-solution display MI. **Bottom Left Panel**. Time evolution of the modulo square of the dissipative soliton in Eq. (14) as obtained from direct numerical integration of Eq. (5), for parameter values $\gamma = 0.01, \nu = 3, \delta = -0.01$. Other parameters are derived from Eq. (16) as $\beta = 0.962424, \alpha = 0.01/3$. **Bottom right panel**. Time evolution of a dissipative soliton of Eq. (5) in presence of a linear force $\epsilon n \psi_n$ of strength $\varepsilon = 0.0002$. Other parameters are fixed as in bottom left panel.

(i) Single dissipative discrete soliton. It can be search in the form

$$\psi_n = \frac{\sinh(\beta)}{\cosh(\beta n)} e^{-i\omega z}. \tag{14}$$

Using the relation

$$\text{sech}(a + \beta) + \text{sech}(a - \beta) = 2\frac{\cosh(a)\cosh(\beta)}{\cosh^2(a)+\sinh^2(\beta)}, \tag{15}$$

we obtain that it is the exact solution of Eq. (5) if

$$\beta = \cosh^{-1}\left(\frac{\gamma}{2\alpha}\right), \omega = -\frac{\gamma}{\alpha}, \mu = 1, \omega = -\nu, \gamma = -\delta. \tag{16}$$

(ii) Nonlinear periodic solution. We assume the ansatz form

$$\psi_n = \frac{\text{sn}(\beta,m)}{\text{cn}(\beta,m)} e^{i\omega z} \text{dn}(\beta n, m). \tag{17}$$

Taking into account the relation for the cnoidal functions

$$\text{dn}(a + \beta) + \text{dn}(a - \beta) = 2\frac{\text{dn}(a)\text{dn}(\beta)}{1-m^2\text{sn}^2(a)\text{sn}^2(\beta)}, \tag{18}$$

we find that the solution parameters should be taken as:

$$\omega = \frac{\delta}{\alpha} = -\nu, \quad \frac{\text{dn}\beta}{\text{cn}^2(\beta)} = \frac{\gamma}{2\alpha}, \quad \gamma = -\delta. \tag{19}$$

(iii) A second type of the nonlinear periodic solution can be obtained from the ansatz

$$\psi_n = \sqrt{m}\, \frac{\text{sn}(\beta,m)}{\text{dn}(\beta,m)} \text{cn}(\beta n) e^{i\omega z}. \tag{20}$$

Taking into account the relation

$$\text{cn}(a + \beta) + \text{cn}(a - \beta) = 2\frac{\text{cn}(a)\text{cn}(\beta)}{m\text{cn}^2(a)\text{sn}^2(\beta)+\text{dn}^2(\beta)}. \tag{21}$$

we find that in this case the parameters must satisfy

$$\omega = \frac{\delta}{\alpha} = -\nu, \quad \frac{\operatorname{cn}\beta}{\operatorname{dn}^2(\beta)} = \frac{\gamma}{2\alpha}, \quad \gamma = -\delta. \qquad (22)$$

We remark that for the periodic solutions the above parameter relations must be complemented with the periodicity condition $\beta N_p = X_p$ where $N_p$ is the number of points per spatial period ($X_p = 2K(m)$ for Eq. (17) and $X_p = 4K(m)$ for Eq. (20), with $K(m)$ the complete elliptic integral of first kind).

The stability of the above exact solutions dissipative can be checked by direct numerical integrations of Eq. (5) taking as initial conditions the exact solutions with a small noise component added in order to accelerate the emergence of eventual instabilities. In the top left panel of Fig.6 we show the time evolution of the periodic dissipative soliton trains in Eqs. (17) and (20). We see that while the $cn$ solution remains stable over a long time, the $dn$ solution display modulational instability at the propagation length $z \approx 200$ out of which two single humps dissipative solitons are created.

Notice from the top right panel of Fig.6 that the $dn$ solution is unstable. This correlates with the fact that it can be seen as an uniform $k = 0$ background with superimposed a plane wave of wavenumber $Q = 0.628$ in correspondence of which the analysis of the previous section predicts instability with a MI gain of $\approx 0.561$. Moreover, we see that out of the instability emerge bright solitons as it is expected for attractive (focusing) nonlinearity.

In the left bottom panel of Fig.6 we show the propagation of the single hump dissipative soliton in Eq. (14) which is the limit of an infinite period ($m \to 1$) of the soliton trains in Eqs. (17) and (20). We see that this soliton is also very stable under long propagation distances. We remark that this soliton can exist only due to the perfect balance between the linear damping ($\delta < 0$) and the nonlinear amplification a condition which can be realized only in the stationary case.

As soon as one deviate from stationarity, as for example is the case when external forces or potentials try to put the soliton in motion, the soliton may become dynamically unstable under time evolution. To investigate this dynamical instability we add a linear potential of the type $\epsilon n\psi_n$ in the right hand side of Eq. (5) which can be implemented in an optical context by a curved optical fiber. The resulting dynamics of the dissipative soliton is depicted in the bottom right panel of Fig.6. We see that, a part for small oscillations, the soliton can survives for a long time the acceleration process without significative changes in its shape. By reducing the strength of the linear potential, pinning phenomena can also become possible, this occurring in the figure at $z \approx 1200$. In this case the onsite symmetric soliton becomes pinned to a lattice site in a state for which the perfect balance between damping and amplification is not realized, this leading to the instability of the state.

## VI. Conclusion

In this chapter we have reviewed some of the linear and nonlinear properties of localized and extended dissipative states of the discrete complex Ginzburg-Landau equation. In particular, we discussed the linear band structure in presence of gain and loss and the existence of onsite and intra-site symmetric discrete solitons for specific and typical parameters. The modulational instability problem of the nonlinear plane waves in the presence of a saturable nonlinearity was also considered and analytical expressions for exact localized and periodic solitons in the (cnoidal waves) derived. It was shown that in the region of the parameter space where the MI gain is positive, generation of solitons and nonlinear periodic wave structures is possible. By taking specific parameters in these regions we found that while discrete soliton and cnoidal waves of cn-type are stable, solutions of dn-types on finite backgrounds are modulationally unstable. We also considered the effect of a linear ramp on a stable localized dissipative soliton and showed that the soliton could propagate under such a disturbance for relatively long distances.